\documentclass[a4paper,11pt]{article}
\pdfoutput=1 %

\usepackage{jcappub} %

\usepackage[T1]{fontenc} %
\usepackage[latin1]{inputenc}
\usepackage[T1]{fontenc}
\usepackage{textcomp}
\usepackage{amsmath}
\usepackage{amsfonts}
\usepackage{amssymb}
\usepackage{makeidx}
\usepackage{graphicx}
\usepackage{xcolor}
\usepackage{ragged2e}
\usepackage{array}
\usepackage{float}
\usepackage{tabularx}
\usepackage{calrsfs}
\usepackage{soul}
\usepackage{siunitx}
\usepackage[symbol]{footmisc}
\usepackage[left=1.00in, right=1.00in, top=1.00in, bottom=1.00in]{geometry}
\usepackage[caption=false]{subfig}
\usepackage{natbib}

\DeclareMathOperator{\Real}{Re}
\DeclareMathOperator{\Imag}{Im}
\DeclareMathOperator{\match}{match}
\title{\boldmath Ringdown Spectroscopy of Rotating Black Holes Pierced by Cosmic Strings}

\author[*,\dagger,a]{Mark Ho-Yeuk Cheung,\note{Corresponding author}\note{These authors contributed equally.}}
\author[\dagger,a]{Levi Wing-Hei Poon,}
\author[a,b]{Adrian Ka-Wai Chung}
\author[a,c,d]{and Tjonnie Guang Feng Li}

\affiliation[a]{Department of Physics, The Chinese University of Hong Kong,\\ Shatin, N.T., Hong Kong}
\affiliation[b]{Theoretical Particle Physics and Cosmology Group,\\ Department of Physics, King's College London, University of London,\\ Strand, London, WC2R 2LS, U.K.}
\affiliation[c]{Institute for Theoretical Physics, KU Leuven,\\ Celestijnenlaan 200D, B-3001 Leuven, Belgium}
\affiliation[d]{Department of Electrical Engineering (ESAT), KU Leuven,\\ Kasteelpark Arenberg 10, B-3001 Leuven, Belgium}

\emailAdd{hycheung@link.cuhk.edu.hk}
\emailAdd{whlpoon@link.cuhk.edu.hk}
\emailAdd{ka-wai.chung@ligo.org}
\emailAdd{tgfli@cuhk.edu.hk}

\abstract{Multiple gauge theories predict the presence of cosmic strings with different mass densities $G\mu/c^2$. 
	We derive an equation governing the perturbations of a rotating black hole pierced by a straight, infinitely long cosmic string along its axis of rotation and calculate the quasinormal-mode frequencies of such a black hole. 
	We then carry out parameter estimation on the first detected gravitational-wave event, GW150914, by hypothesizing that there is a string piercing through the remnant, yielding a constraint of $G\mu/c^2 <3.8\times 10^{-3}$ at the 90\% confidence interval with a comparable Bayes factor with an analysis for a Kerr black hole without a string.
	In contrast to existing studies which focus on the mutual intersection of cosmic strings, or the cosmic string network, our work focuses on the intersection of a cosmic string with a black hole, with characteristics which can be identified in binary coalescence signals.}

\begin{document}
\maketitle
\flushbottom

\section{Introduction}
The detection of gravitational wave events by the Advanced LIGO and VIRGO detectors \cite{LIGO_01,LIGO_02,LIGO_03,LIGO_04,LIGO_05,LIGO_06,LIGO_08,LIGO_09} has opened a new window in terms of astrophysical observations.
Not only is it now possible to detect black hole and neutron star merger events, but also to infer the properties of the merging objects by analysing their gravitational-wave signals. 
In particular, analysing the quasinormal-mode frequencies (QNMFs) \cite{nollert1999quasinormal,kokkotas1999quasi,berti2009quasinormal,Konoplya:2011qq} of the decaying waves at the final ringdown stage of a binary black-hole merger event could help infer the different parameters that characterize the remnant black hole. 
A cosmic string hair of black holes might also be probed in the same way.

Cosmic strings are hypothetical topological defects predicted by some gauge theories and could have formed during cosmological phase transition in the early universe \cite{Vilenkin2000,Copeland2010}.
Efforts have been made to search for cosmic strings directly, or to constrain their mass density $\mu$ by analyzing the cosmic microwave background \cite{Pogosian1999,Ade2014,Lizarraga2014,Lazanu2015,Lizarraga2016}, stochastic gravitational-wave background \cite{Vilenkin1981,Hogan1984,Brandenberger1986,Accetta1989, Ringeval2017,Blanco2018,blancopillado2019energyconservation}, by considering their lensing effects \cite{De1997,Mack2007,Thomas2009,Jung2018} and bursts of gravitational wave produced by cosmic string cusps or kinks \cite{Damour2001, Abbott2019}.
The most stringent constraint put on the cosmic string network is $G\mu/c^2 <1.5\times 10^{-11}$ \cite{Blanco2018}, while it is estimated that this constraint can be pushed to $G\mu/c^2 < 10^{-17}$ with LISA \cite{auclair2019probing}, where $G$ is the gravitational constant. 

Other than existing within its own network, cosmic strings might also be found piercing through black holes.
Such a configuration could form when a black hole with non-zero magnetic charge cools below the transition temperature \cite{Aryal1986}, or when a primordial black hole forms around a string \cite{Vilenkin2018}.
It has been shown that cosmic strings can be stably supported by black hole horizons, giving rise to long range black hole hair \cite{aryal1986cosmic,achucarro1995abelian,bonjour1999vortices,dehghani2002vortex,dehghani2002abelian,ghezelbash2002vortices,ghezelbash2002abelian,gregory2013rotating,nakonieczny2013abelian,gregory2014vortex,Kubizvnak2015}, possibly serving as a counter-example to the no hair theorem \cite{chrusciel1994no,bekenstein1996black}.
If the black hole is rotating, the string will align with the axis of rotation of the black hole at equilibrium \cite{Kubizvnak2015}, hence we call such a black hole a Kerr-string black hole.
The QNMFs of a non-rotating Schwarzschild-string black hole have been calculated in Refs.~\cite{chen2008quasinormal,SINI_2009}, but not for the case of a Kerr-string black hole with non-zero angular momentum.

In this paper, we will show that cosmic string hair of a Kerr black hole would affect its ringdown waveforms, providing a way to search for cosmic strings by gravitational-wave detection and ringdown-spectroscopy analysis.
The QNMFs of a Kerr-string black hole are first calculated by solving a modified Teukolsky equation that takes the effect of the cosmic string into account. 
Then, we will constrain the mass density of cosmic strings piercing through the remnant black hole formed in the merger event GW150914 detected by LIGO \cite{LIGO_01} with its ringdown signal, as well as discuss the degeneracies between different parameters in our analysis.

Throughout the remainder of this paper, we will adopt the $(+,-,-,-)$ convention and units where $G=c=1$ will be used. However, to remain consistent with the literature, we will keep the $G$ in front of $\mu$.

\section{The Perturbation Equation}
The metric of a Kerr-string black hole, consisting of a black hole with mass $ M $ and specific angular momentum $ a $ pierced by a cosmic string of infinite length along the axis of rotation with dimensionless mass density $ G\mu\ll 1$, is given by \cite{CS}
\begin{equation}\label{metric}
    ds^2=\frac{\Delta\Sigma}{\Gamma}dt^2-\frac{\Gamma\sin^2\theta}{\Sigma}\left(b\ d\phi-\frac{2aMr}{\Gamma}dt\right)^2-\Sigma\left(\frac{dr^2}{\Delta}+d\theta^2\right),
\end{equation}
where $b=1-4G\mu$, $\Sigma=r^2+a^2\cos^2\theta$, $\Delta=r^2+a^2-2Mr$ and  $\Gamma=(r^2+a^2)^2-\Delta a^2\sin^2\theta$. 
This metric can be obtained by introducing an azimuthal angular deficit of $8\pi G\mu$ on $\phi$ to the Kerr metric, and it remains a Petrov type D vacuum metric, suggesting that the Teukolsky formalism of treating black hole perturbation could be applied to the system.

Locally, Eq.~\eqref{metric} is equal to the Kerr metric with $\phi$ rescaled to $b\phi$, so the perturbation equation for Eq.~\eqref{metric} can be obtained by repeating Teukolsky's computation \cite{TE} with the following null tetrad:
\begin{align}
l^\mu &= \dfrac{1}{\Delta}\left\langle r^2+a^2,\Delta,0,\frac{a}{b}\right\rangle,\\
n^\mu &= \frac{1}{2\Sigma}\left\langle r^2+a^2,-\Delta,0,\dfrac{a}{b}\right\rangle,\\
m^\mu &= \frac{1}{\sqrt{2}(r+ia\cos\theta)}\left\langle ia\sin\theta,0,1,\frac{i}{b\sin\theta}\right\rangle,\\
\overline{m}^\mu &= \frac{1}{\sqrt{2}(r-ia\cos\theta)}\left\langle-ia\sin\theta,0,1,\frac{-i}{b\sin\theta}\right\rangle.
\end{align}
Then, the master equation for black holes pierced by cosmic strings is derived to be
\begin{equation}\label{newTE}
    \begin{split}
    \left[\frac{(r^2+a^2)^2}{\Delta}-a^2 \sin^2\theta\right]\dfrac{\partial^2 \Psi}{\partial t^2}+\frac{4Mar}{\Delta}\dfrac{\partial^2 \Psi}{\partial t\partial\phi}+\frac{1}{b^2}\left[\frac{a^2}{\Delta}-\frac{1}{\sin^2\theta}\right]\dfrac{\partial^2 \Psi}{\partial\phi^2}&\\
    -\Delta^{-s}\dfrac{\partial}{\partial r}\left(\Delta^{s+1}\dfrac{\partial \Psi}{\partial r}\right)
    -\frac{1}{\sin\theta}\dfrac{\partial}{\partial \theta} \left(\sin\theta\dfrac{\partial \Psi}{\partial \theta}\right)
    -\frac{2s}{b}\left[\frac{a(r-M)}{\Delta}+i\frac{\cos\theta}{\sin^2\theta}\right]\dfrac{\partial \Psi}{\partial\phi}&\\
    -2s\left[\frac{M(r^2-a^2)}{\Delta}-r-ia\cos\theta\right]\dfrac{\partial\Psi}{\partial t}+(s^2\cot^2\theta-s)\Psi
    &=4\pi \Sigma T.
    \end{split}
\end{equation}
This is equivalent to the original Teukolsky equation with $\phi$ replaced by $b\phi$.
\section{Quasinormal Mode Frequencies} \label{Sec:QNM}
By considering a vacuum perturbation (i.e., $T=0$), we can compute the QNMFs using Leaver's method of continued fraction \cite{Leaver,berti2009quasinormal}.
Analogous to Teukolsky's computations \cite{TE}, by putting $\Psi=e^{i(m\phi-\omega t)}R(r)S(\theta)$, Eq.~\eqref{newTE} is separable into a radial equation and an angular equation:
\begin{equation}\label{radial}
    \Delta^{-s}\dfrac{d}{dr}\left(\Delta^{s+1}\dfrac{dR}{dr}\right)
    +\left[\dfrac{K^2-2is(r-M)K}{\Delta}+4is\omega r-\lambda\right]R=0,
\end{equation}
\begin{equation}\label{polar}
    \dfrac{1}{\sin\theta}\dfrac{d}{d\theta}\left(\sin\theta\dfrac{dS}{d\theta}\right)   +\left[\vphantom{\dfrac{m/b-s\cos\theta}{\sin\theta}}\left(a\omega\cos\theta-s\right)^2\right.
    \left.-\left(\dfrac{m/b-s\cos\theta}{\sin\theta}\right)^2-s^2+s-A\right]S=0,
\end{equation}
where $K=(r^2+a^2)\omega-am/b$, $\lambda=A+a^2\omega^2-2am\omega/b$ and $A$ is a separation constant.
Effectively, the inclusion of a cosmic string amounts to changing $m$ to $m/b$ in Eq.~\eqref{radial} and Eq.~\eqref{polar}, the QNMFs can hence be computed by changing the value of $m$ in Leaver's algorithm.
\begin{figure}[h]
    \centering
    \includegraphics[width=0.7\textwidth]{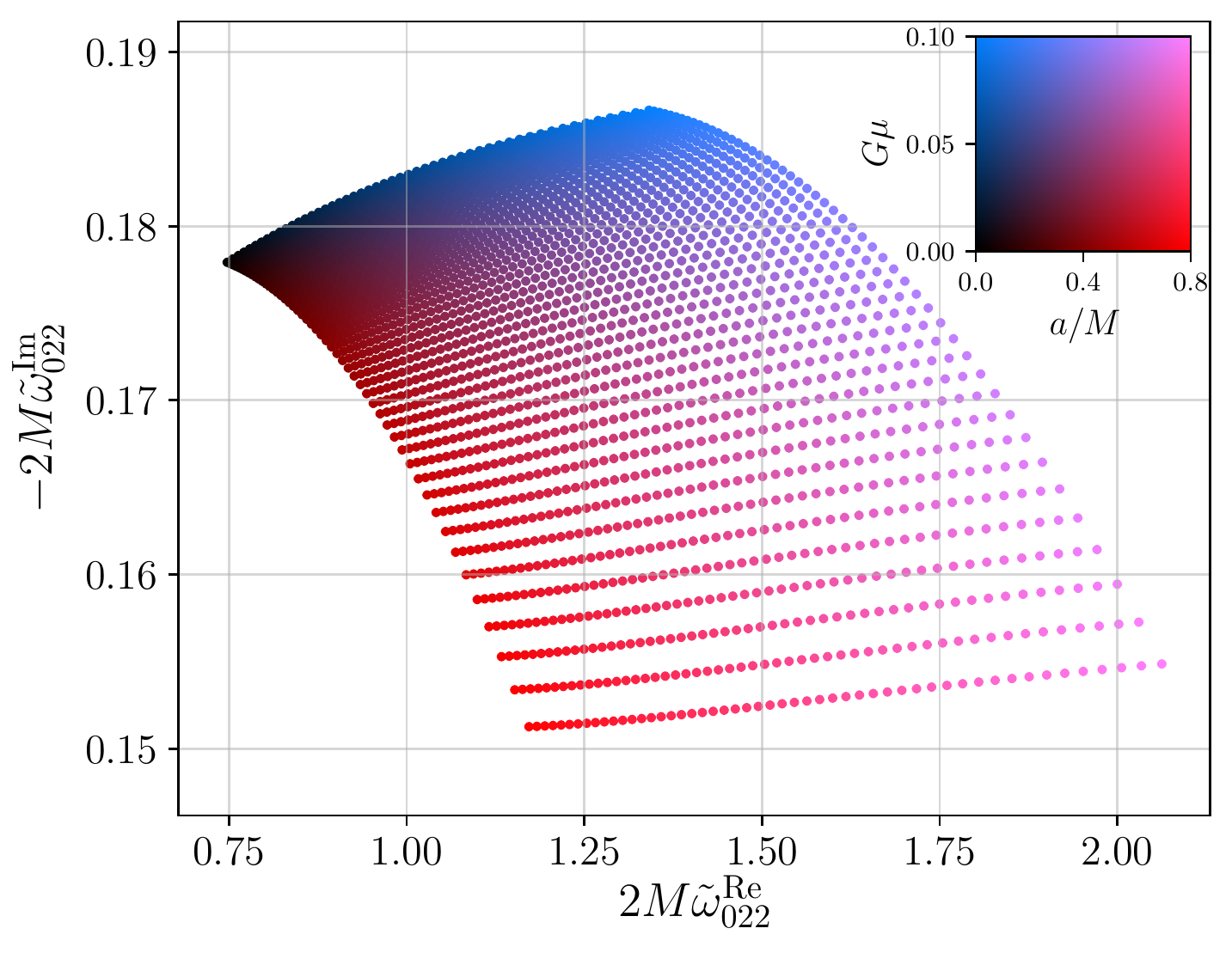}
    \caption{The real and imaginary parts of the $ (n,l,m)=(0,2,2) $ overtone QNMFs for $0\le G\mu\le 0.1$ and $0\le a/M\le 0.8$.
    Blueness and redness represent the values of $G \mu$ and $a$ respectively.
    The dots are evenly spaced in $G\mu$ and $a$.
	This shows that increasing $G\mu $ increases the magnitudes of both the real and imaginary parts of the QNMFs $ \tilde{\omega}_{022} $.}
    \label{fig:QNMF}
\end{figure}

Figure~\ref{fig:QNMF} shows the dependence of the $ (n,l,m)=(0,2,2) $ overtone frequency on $G\mu$ at various values of the spin parameter $a/M$.
A bluer hue represents a larger value of $G\mu$ and a redder hue represents a larger value of $a$.
Both the frequency and the decay rate of gravitational wave emissions increase with the mass density of the cosmic string.
Notably, an increasing spin and an increasing $G\mu$ drive the $(0,2,2)$ overtone QNMFs in different directions.
This suggests that parameter estimation through ringdown analysis is possible.
Moreover, none of the common alternative gravitational theories that would affect the QNMFs increase both $|\tilde{\omega}^{\text{Im}}|$ and $|\tilde{\omega}^{\text{Re}}|$ together  \cite{cardoso2009perturbations,blazquez2019quasinormal,Chung_2019,McManus_2019,Bao_2019}, so parameter estimation on $G\mu$ done with the QNMFs will not be degenerate with these theories.

Interesting trends are found in the QNMFs of black holes with cosmic strings.
In the Schwarzschild case ($a = 0$), the separation constant is given by $A_{lm}=l(l+1)-s(s+1)$ \cite{TE}.
With the effect of a cosmic string included, for modes with $n=0$ and $l=m=2,3,4$, it is noticed that the separation constant is modified to $A_{lm}=(l/b)(l/b+1)-s(s+1)$ when $ a=0 $ and that  $\omega^\text{Re}(a,G\mu)\approx\omega^\text{Re}(a,0)/b$.
Such a relation is not present in modes with $l\neq m$.

The QNMFs for each $ nlm $ overtone was fitted to a quadratic function in $G\mu$:
\begin{equation}
    \tilde{\omega}_{nlm}(a,G\mu)=\tilde{\omega}_{nlm}(a,0)+\tilde{\omega}_{nlm}(0,0)[\tilde{c}_1(a)G\mu+\tilde{c}_2(a)(G\mu)^2]. 
\end{equation}
The real and imaginary parts of $\tilde{c}_1(a)$ and $\tilde{c}_2(a)$ were then fitted to the form $k_1-k_2(1-a/M)^{k_3}$ as motivated by Ref.~\cite{echeverria1989gravitational}.
For the $(0,2,2)$ mode, which is the dominant mode in GW150914, the maximum error of the fit over the range $0\le G\mu\le0.1$, $0\le a/M\le0.95$ in the real and imaginary parts of the QNMFs is about 0.8\% and 1.4\% respectively.
Table~\ref{tab:coeff} shows the numerical values of the coefficients for this mode.
\begin{table}[h]
    \centering
    \begin{tabularx}{\columnwidth}{>{\centering\arraybackslash}X>{\centering\arraybackslash}X>{\centering\arraybackslash}X>{\centering\arraybackslash}X}
    \hline\hline
    $f(a)$ &$k_1$ &$k_2$&$k_3$ \\
     \hline
    $\Real\{\tilde{c}_1(a)\}$ & $10.1$ & $5.73$ & $0.223$ \\
     $\Imag\{\tilde{c}_1(a)\} $ &$2.34$ & $1.43$ & $0.289$  \\
    $\Real\{\tilde{c}_2(a)\}$ &$108$ & $77.7$ & $0.180$   \\
     $\Imag\{\tilde{c}_2(a)\}$ &$60.7$ & $53.4$ & $0.0503$ \\
    \hline\hline
    \end{tabularx}
    \caption{Fitted coefficients for $f(a)=k_1-k_2(1-a/M)^{k_3}$ for different $f(a)$ of the $(0,2,2)$ mode, where $ f(a) $ are the real and imaginary parts of coefficients appearing in a quadratic of the QNMFs to $ G\mu $.}
    \label{tab:coeff}
\end{table}
\begin{figure}
    \centering
    \includegraphics[width=0.7\textwidth]{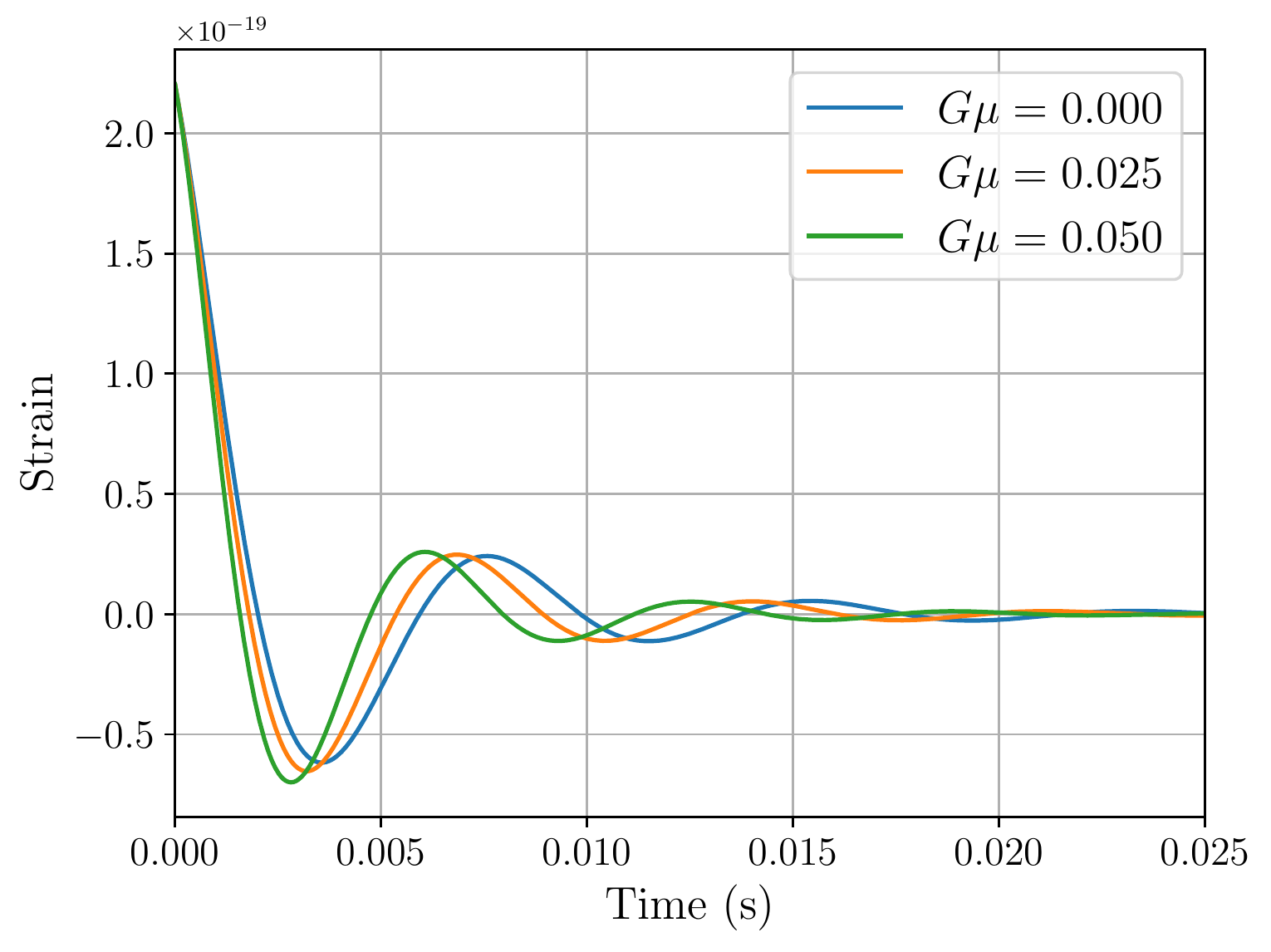}
    \caption{Ringdown waveforms generated by Kerr-string black holes with different $G\mu$. Only the $(0,2,2)$ and $(1,2,2)$ modes are used in the generation of the waveforms. The frequency of QNM oscillation increases with $G\mu$ while the lifetime decreases slightly.  }
    \label{fig:waveforms}
\end{figure}

Figure~\ref{fig:waveforms} plots the ringdown waveforms of a Kerr-string black holes of different $G\mu$.
The waveforms are assumed to be generated from a black hole merger event where the $(n,l,m)=(0,2,2)$ and $(1,2,2)$ modes are dominant in the ringdown signal, and the mass of the final black hole $M_f$ and final spin parameter $a_f$ are the same for all the waveforms. 
Using two or more modes instead of one, we can ensure that our waveform has four degrees of freedom in its QNMFs, so it can carry information about all three parameters $M, a$ and $G\mu$.
Altering $G\mu$ would induce a change in the shape of the waveform, and the higher the value of $G\mu$, the more the waveform deviates from that from a system without a cosmic string.

Before making use of the QNMFs fits to constrain $G\mu$ with real data, it would be insightful to look into possible degeneracies between $M_f$ or $a_f$ with $G\mu$.
 We do this by considering the match between two waveforms, one with fixed parameters and $G\mu=0$, the other with non-zero $G\mu$ and varying $M_f$ or $a_f$.
The match between waveforms is defined by 
\begin{equation}
\match[h_1,h_2] = \dfrac{\langle h_1 | h_2\rangle}{\sqrt{\langle h_1 | h_1\rangle \langle h_2 | h_2\rangle}},
\end{equation}
with
\begin{align}
&h_1 \equiv h_1(M_{f1},a_{f1},G\mu_1;f), \\
&h_2 \equiv h_2(M_{f2},a_{f2},G\mu_2;f),
\end{align}
and
\begin{equation}
\langle h_1| h_2\rangle= \int^{f_\mathit{\rm high}}_{f_\mathit{\rm low}} \dfrac{h^*_1(f) \, h_2(f)}{S_n(f)} df,    
\end{equation}
where $h_1(M_{f1},a_{f1},G\mu_1;f)$ and $h_2(M_{f2},a_{f2},G\mu_2;f)$ correspond to ringdown waveforms in the frequency domain with different parameters $M_f$, $a_f$ and $G\mu$, $S_n(f)$ is the power-spectral density (PSD) of LIGO, and $^*$ denotes complex conjugation. 
The duration of the ringdown signals in the time domain is set to be $\SI{0.030}{\second}$, and their match in the frequency domain is calculated with \textsc{PyCBC} \cite{pycbc}, with $f_\mathit{\rm low} = \SI{30}{\Hz}$ and $f_\mathit{\rm high}$ equal to the highest frequency attained by the ringdown waveform.

\begin{figure}[h]\label{fig:match}
\centering
\subfloat[\label{sfig:contour_plot_mass}]{%
  \includegraphics[width=0.7\textwidth]{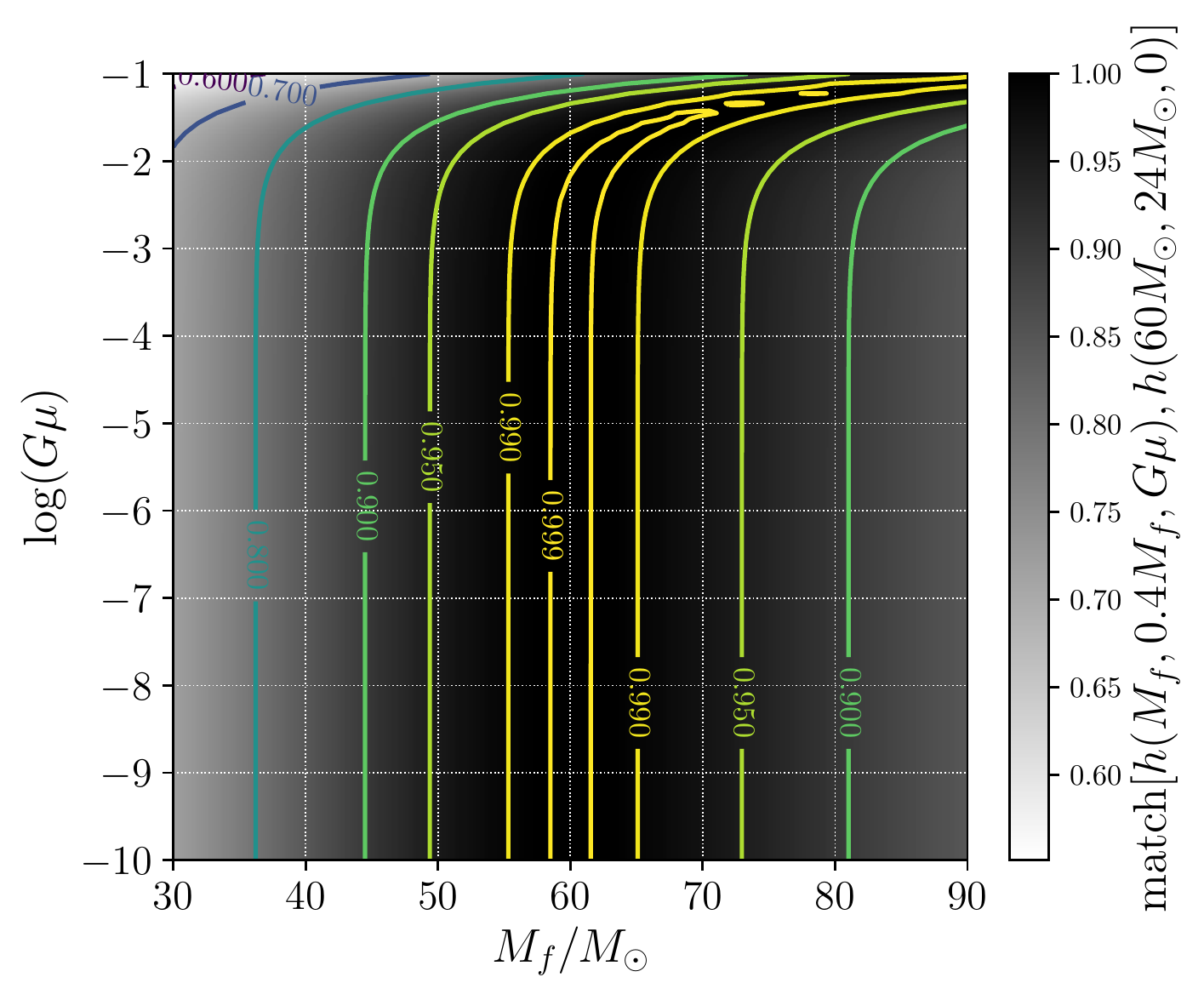}%
}\hfill
\subfloat[\label{sfig:contour_plot_spin}]{%
  \includegraphics[width=0.7\textwidth]{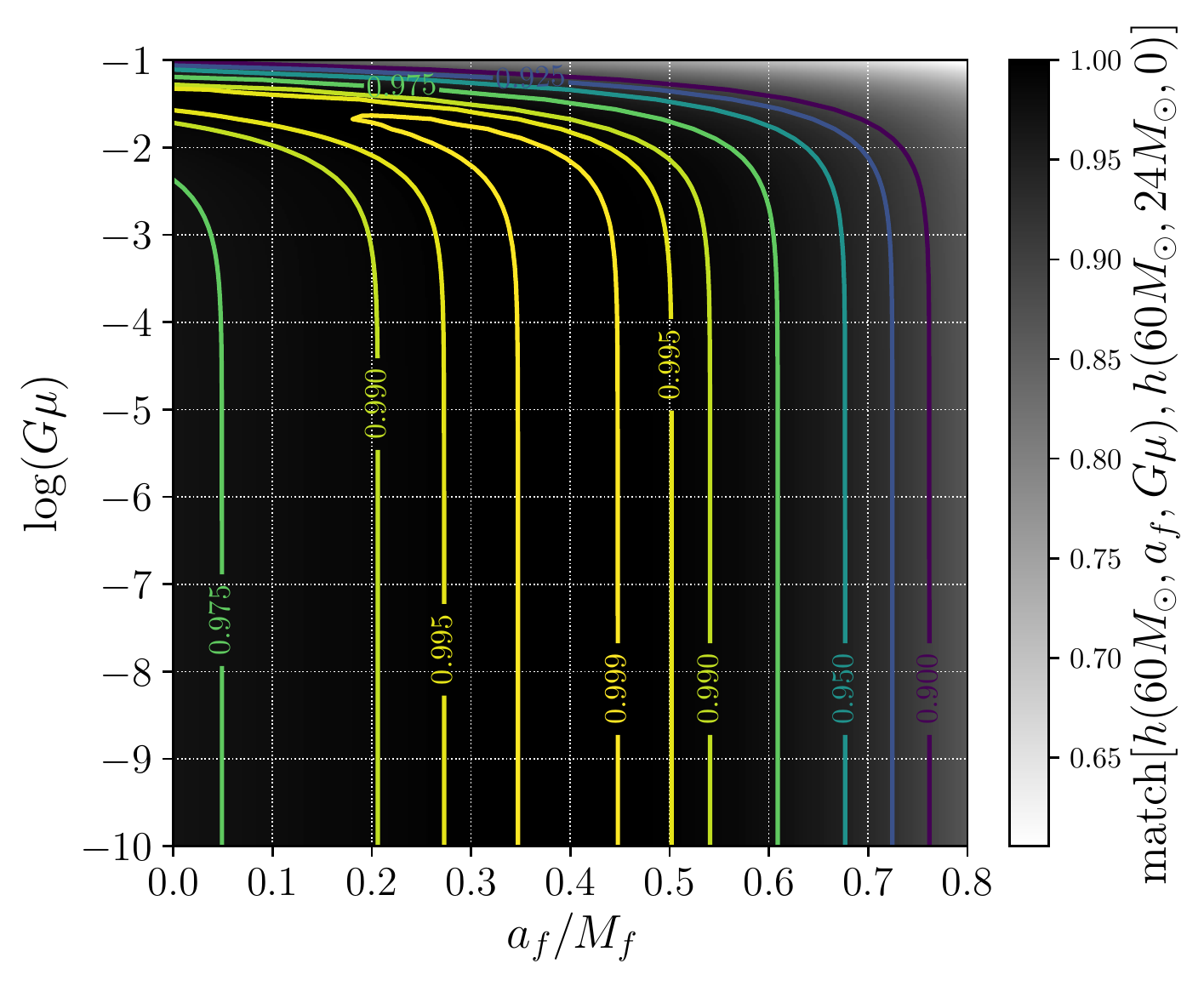}%
}
\caption{
Contour plots of match between waveforms with different parameters to test the degeneracy of $G\mu$ with $M_f$ and $a_f$.
\protect\subref{sfig:contour_plot_mass}: Match between a fixed string-less waveform with $M_f$ fixed at $60M_\odot$ and one with variable $G\mu$ and $M_f$ (both with $a_f/M_f$ fixed at $0.4$). 
Degeneracy is observed for $\log G\mu > -3$, where a stringless waveform will look like a string-carrying waveform with a higher mass.
\protect\subref{sfig:contour_plot_spin}: Match between a fixed string-less waveform with $a_f/M_f=0.4$ and one with variable $G\mu$ and $a_f$ (both with $M_f$ fixed at $60M_\odot$). 
Similar to \protect\subref{sfig:contour_plot_mass}, there is degeneracy observed for $\log G\mu > -3$, but with a string-less waveform looking like a string-carrying waveform with less spin.
}
\label{contour_plots}
\end{figure}
The match between the waveforms $h(M_f,0.4M_f,G\mu)$ and $h(60M_\odot,24M_\odot,0)$ is plotted in Figure~\ref{sfig:contour_plot_mass}.
By varying $M_f$ and $G\mu$, it is found that the two parameters are degenerate in the $\log(G\mu)>-3$ regime.
Nonetheless, for sufficiently high values of the match, the majority of the contour area is still centered at $M_f=60M_\odot$ and terminates at some high value of $\log(G\mu)$, meaning that it is still possible to constrain $G\mu$ by matched filtering without affecting too significantly the parameter estimation of $M_f$.
Similarly, Figure~\ref{sfig:contour_plot_spin} test the degeneracy of $G\mu$ with $a_f$. 
Again, there is degeneracy in the high $\log(G\mu)$ regime, which will cause a string-less waveform to look like a string-carrying waveform with less spin.
This will cause the constrain on $G\mu$ to be less tight.

\section{Results}
The uncertainty in the measurement of $M_f$ and $a_f$ of the remnant black holes of detected gravitational-wave events have left room for the possible existence of black hole hair \cite{Konoplya:2016pmh}.
If GW150914 were a merger of primordial black holes (as suggested by \cite{sasaki2016primordial,blinnikov2016solving}), it might be possible for at least one of the black holes in the binary to hold cosmic string hair.
When this Kerr-string black hole merges with the other black hole, the remnant will also be a Kerr-string black hole and it will emit a ringdown signal characterized by the parameters $M_f, a_f$ and $G\mu$.
Thus, we can hypothesize that the ringdown signal of GW150914 comes from a Kerr-string black hole and constrain $G\mu$ of the hypothetical string piercing through it.
Such parameter estimation is done with the QNMFs of the Kerr-string black hole and with the \textsc{pyRing} pipeline introduced in Ref.~\cite{code}. 

As mentioned earlier, we will have to use at least two ringdown modes to estimate the three parameters $M, a$ and $G\mu$.
We might have chosen to use the $(0,2,2)$ and $(0,2,1)$ modes in theory due to them being the two most excited modes of GW150914 \cite{Kamaretsos_2012}, but the large $G\mu$ $(0,2,1)$ QNMFs have values too close to those of low $G\mu$ $(0,2,2)$, which will cause the pipeline to falsely recognize high $G\mu$ waveforms in the data.
Therefore, the $(0,2,2)$ and $(1,2,2)$ ringdown modes are used.
The prior distribution for $\log(G\mu)$ and start time are respectively set to be a uniform distribution within $[-10,-1]$ and $ [10,20]M $ \cite{Bhagwat2018, Carullo_prior}, where $M=68M_\odot$ is the reported median of the final mass in Ref.~\cite{LIGO_07}. 
We follow the choices of Ref.~\cite{code} for the prior distributions of other parameters as well as the treatment of noise of the LIGO data.

\begin{figure}[!htbp]
	\centering
	\includegraphics[width=0.7\textwidth]{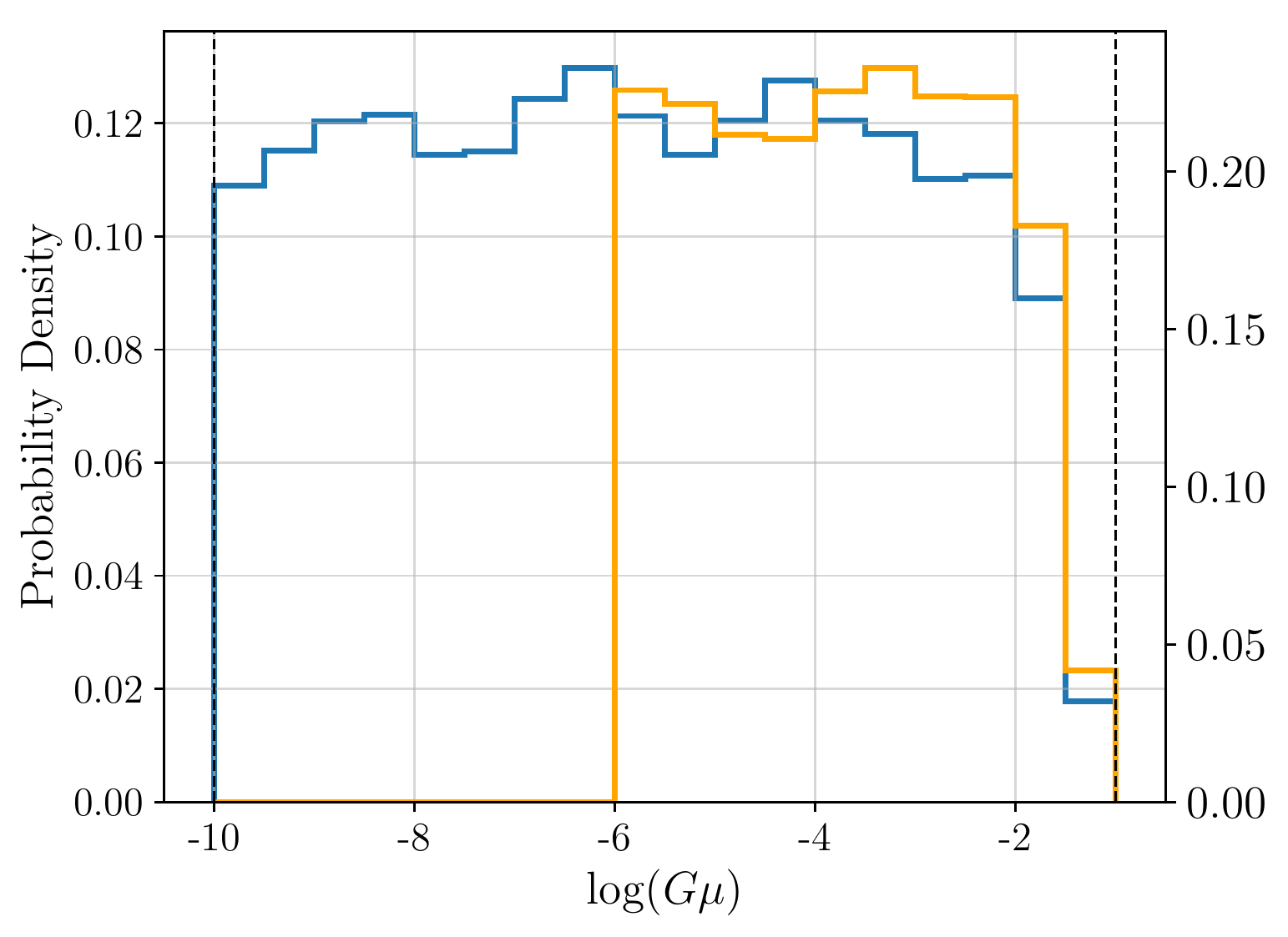}
	\caption{The posterior distributions of $\log(G\mu)$ for the GW150914 event with prior bounds $ -10\le\log(G\mu)\le -1 $ (blue, left scale) and $ -6\le\log(G\mu)\le-1 $ (orange, right scale). The shape of the distribution resembles that of a step function. There is a sharp drop near $\log(G\mu)=-2$ for both ranges, indicating that this constraint can be obtained independently from the prior range. We obtain a constraint of $G\mu<3.8\times 10^{-3}$ at the 90\% confidence level for the $ [-10,-1] $ range. The prior bounds are marked with dotted lines.}
	\label{fig:real_pos}
\end{figure}

Figure~\ref{fig:real_pos} shows the posterior distribution of $ \log(G\mu) $.
The relatively sharp drop near $ \log(G\mu)=-2 $ indicates that our analysis can rule out high-$ G\mu $ waveforms in the GW150914 signal.
A constraint of $ G\mu<3.8\times 10^{-3} $ at the 90\% confidence level is obtained.
We also included a posterior distribution obtained from another prior range, $[-6,-1]$ (in orange), to show that the constraint is not significantly affected by the prior bounds.

\begin{figure}
    \centering
	\includegraphics[width=0.7\textwidth]{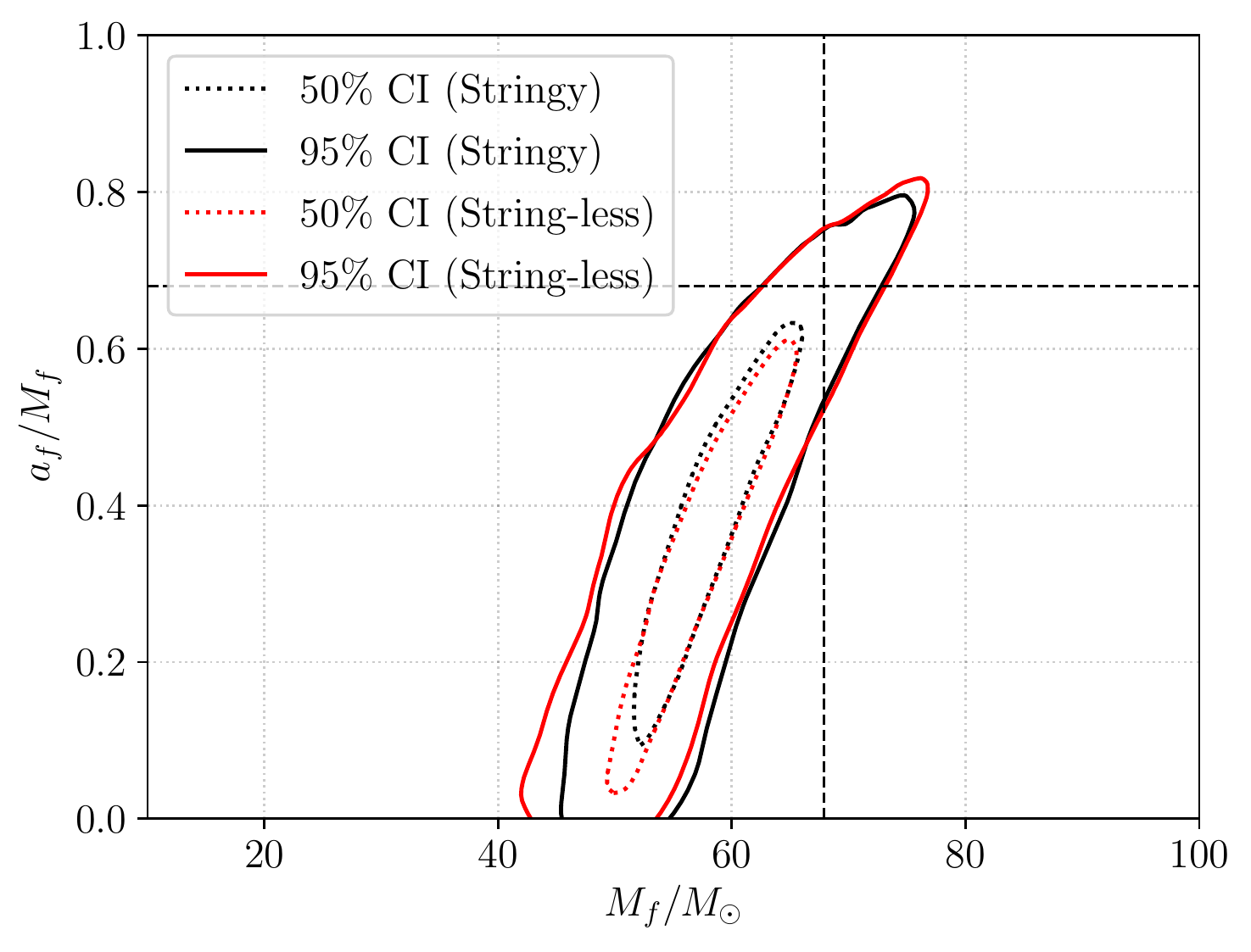}
    \caption{Contour plots of the final mass $M_f$ and spin parameter $a_f/M_f$ in the generated posterior data set for GW150914 for analysis with (in black) and without (in red) an additional parameter estimation on $ G\mu $.
    The vertical and horizontal black dotted lines represent the published median values for the detector-frame final mass $M_f$ and final spin  $a_f/M_f$ of the GW150914 event given in Ref.~\cite{abbott2016properties}.
	The shape of the two plots are similar.}
    \label{fig:real_Mfaf}
\end{figure}

\begin{figure}[!htbp]
    \centering
    \includegraphics[width=0.7\textwidth]{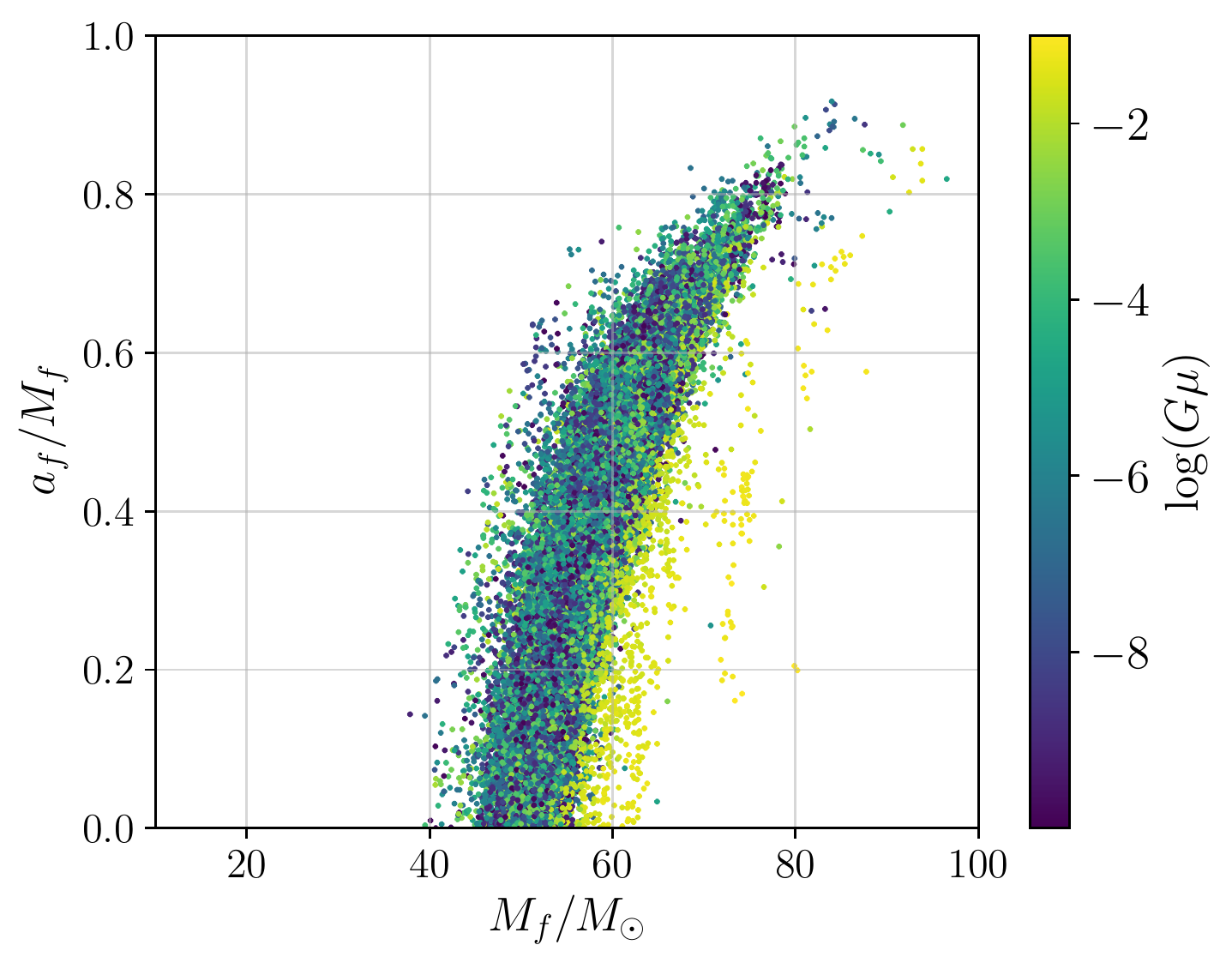}
    \caption{Posterior plot of all three parameters $M_f, a_f$ and $G\mu$. The yellow hue corresponds to points with higher $G\mu$, and they are located more towards the high $M_f$ and low $a_f$ side, agreeing with our waveform degeneracy analysis in Sec.~\ref{Sec:QNM}. }
    \label{fig:Degeneracy}
\end{figure}

Figure~\ref{fig:real_Mfaf} shows the contour plot of the final spin parameter $a_f/M_f$ and final mass $M_f$, with the median values estimated by a full IMR analysis \cite{LIGO_01} marked by two black straight dotted lines.
When compared to the same graph for an analysis without the parameter estimation of $G\mu$ , it can be seen that the plots have generally the same shape.
Figure~\ref{fig:Degeneracy} plots explicitly the three dimensional posterior of $M_f$, $a_f$ and $G\mu$. 
It can be clearly seen that the higher $G\mu$ points lie to the higher $M_f$ and smaller $a_f$ side, agreeing with our waveform degeneracy analysis in Sec.~\ref{Sec:QNM}.

\begin{figure}[!htbp]
    \centering
    \includegraphics[width=0.7\textwidth]{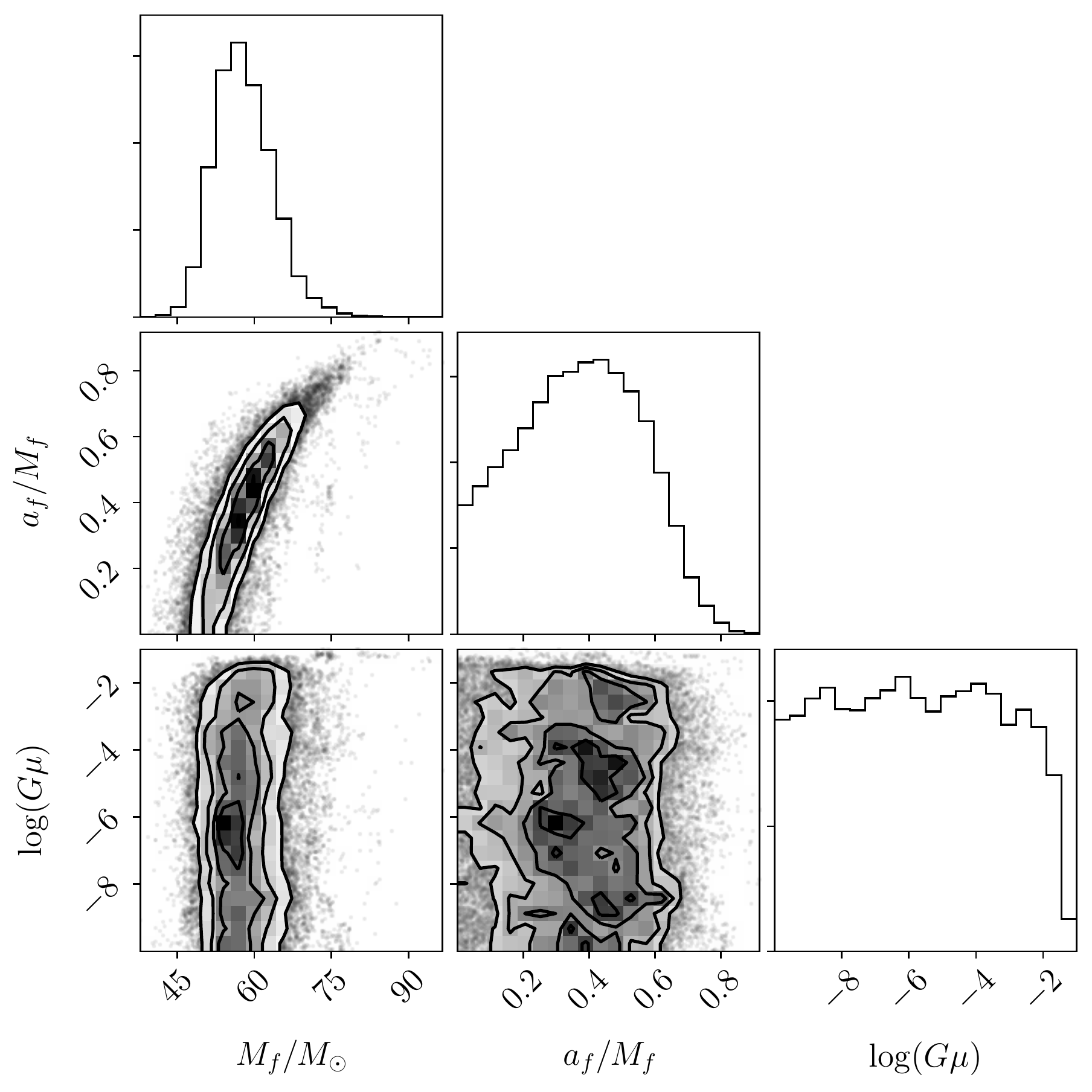}
    \caption{A corner plot showing the posterior distribution of $M_f$, $a_f$ and $\log(G\mu)$. Unlike the contour plot for $a_f$ and $M_f$, no clear trend lines can be observed in the contour plots of $\log(G\mu)$ versus $a_f$ and $M_f$.}
    \label{fig:corner}
\end{figure}

Figure~\ref{fig:corner} shows a corner plot for $M_f$, $a_f$ and $\log(G\mu)$. 
As evident from the absence of clear trend lines in the contour plots of $\log(G\mu)$ against $M_f$ or $a_f$, the degeneracy between the effects of $G\mu$ and those of $M_f$ or $a_f$ are not too significant.
The degeneracy shown in Figure~\ref{fig:match} is stronger than that observed in Figure~\ref{fig:corner} because we pinned one of the parameters $a_f$ or $M_f$ in the former;
the apparent degeneracy is reduced when we include all parameters.

This three-parameter analysis gives a Bayes factor of $ \log B=59.3 $.
When compared with $ \log B=59.1 $ for an analysis with only $M_f$ and $a_f$, this suggests that the two models are similar in plausibility. 

\section{Discussion and Conclusions}
\subsection{Assumptions Made}
For the analysis using QNMFs obtained from Eq.~\eqref{newTE} to be valid, three conditions need to be satisfied:
\begin{enumerate}
    \item \label{itm:precession}The spins of the black holes during the inspiral phase are aligned with the orbital angular momentum vector.
    \item \label{itm:infinite}The string is infinitely long and straight after the merger.
    \item \label{itm:axis}The string lies on the axis of rotation of the final black hole.
\end{enumerate}

(\ref{itm:precession}) is necessary for (\ref{itm:infinite}) and (\ref{itm:axis}):
If (\ref{itm:precession}) is not true, the spin of the Kerr-string black hole will precess, so the string will likely curl around itself, violating (\ref{itm:infinite}).
Moreover, if the spins of the binary black holes are misaligned, the spin of the final black hole might also be misaligned with the string, violating (\ref{itm:axis}). 
A rotating black hole pierced by a cosmic string would gradually approach an equilibrium position where the string aligns with the rotational axis of the black hole \cite{Kubizvnak2015}.
We assumed that the time scale for the approach to this equilibrium is much shorter than the time scale for merger and ringdown so that the ringdown signal is not contaminated by the signal from this stabilization.

In reality, as disturbances travel at a finite speed on the string, even if the spins of the binary black holes and the total angular momentum of the black hole merger are aligned perfectly, the string will spiral with the black holes during the inspiral phase.
However, as the orbital radius of the black holes decreases, the string will steadily approach its final configuration, in which it passes through the center of mass of the system (i.e., the center of the final black hole) and aligns with the system's angular momentum, so it could stabilizes soon after the merger.
In that case, assumptions (\ref{itm:infinite}) and (\ref{itm:axis}) can be satisfied locally and approximately.

When a Kerr-string system is perturbed, the string would also vibrate along with the black hole.
The gravitational radiation of a vibrating wiggly string has a radiation power proportional to $(G\mu)^2$ \cite{Vilenkin2000}, but the change in power induced by the conical deficit of the Kerr-String metric is proportional to $G\mu$, so it is expected that the shift of the QNMFs are contributed more by the conical deficit itself than the vibration of the string.

As a side note, our method also serves as a way to look for cosmic strings that are piercing through supermassive black holes (SMBHs) in extreme mass ratio inspiral (EMRI) events.
In such events, the orbiting stellar mass black hole can be treated as a perturbation that induces negligible back reaction to the metric of the SMBH.
Thus, if the SMBH of the EMRI binary is pierced by a cosmic string, disturbances to the string away from its equilibrium orientation can be neglected, so the string always stays straight and aligns with the axis of rotation of the SMBH, fitting our assumptions on the Kerr-string system.
The QNMs that we calculated in this paper would then be applicable, and it would be straight forward to repeat our analysis on these systems with the methods outlined in the previous sections.
Moreover, as the geodesics of a Kerr-string metric is different from the string-less case, it is expected that the inspiral waveform of such EMRI events can also be used to constrain $G\mu$.

\subsection{Comparison with Existing Results}
Unlike our constraints, methods based on analysis of the cosmic microwave background or the stochastic gravitational wave background apply to entire cosmic string networks.
The effects of the cosmic string network on the background spectrum are model-dependent since one must take the spatial arrangement and temporal evolution of the cosmic strings into account.
Hence, the constraint so obtained will depend on the precise model used to simulate the cosmic string network (see, for example, Ref.~\cite{Ade2014}).

Our methods are more similar to searches for gravitational wave bursts or signs of lensing from cosmic strings in that these are methods of direct search.
The constraint obtained only applies to a single event and a global constraint can only be obtained given models of event rates and arrangements of cosmic strings.

\subsection{Conclusion}
In conclusion, we showed that it is possible to constrain the mass density of cosmic strings piercing rotating black holes by analyzing its ringdown signal.
Although the constraint on the GW150914 event is less stringent than those obtained in Ref.~\cite{Blanco2018}, our work serves as a new way to constrain an individual cosmic string instead of the whole cosmic string network.

\section*{Acknowledgements}

This research has made use of data, software and/or web tools obtained from the Gravitational Wave Open Science Center (https://www.gw-openscience.org), a service of LIGO Laboratory, the LIGO Scientific Collaboration and the Virgo Collaboration \cite{Vallisneri_2015}. LIGO is funded by the U.S. National Science Foundation. Virgo is funded by the French Centre National de Recherche Scientifique (CNRS), the Italian Istituto Nazionale della Fisica Nucleare (INFN) and the Dutch Nikhef, with contributions by Polish and Hungarian institutes.

The authors are indebted to valuable discussion among the Testing General Relativity (TGR) working group of LIGO. 
The authors are grateful to Gregorio Carullo, Xavier Siemens and Mairi Sakellariadou and other members from the Theoretical Particle Physics and Cosmology Group at King's College London for their insightful comments on the manuscript. 
A.K.W.C. is supported by the Hong Kong Scholarship For Excellence Scheme (HKSES). 
The work described in this paper was partially supported by grants from the Research Grants Council of the Hong Kong (Project No. CUHK 24304317), The Croucher Foundation of Hong Kong, and the Research Committee of the Chinese University of Hong Kong.
This document contains a report number of KCL-PH-TH/2020-05.
\bibliographystyle{unsrtnat}
\bibliography{Kerr-string_ringdown}
\end{document}